\documentclass[prl,nofootinbib,floatfix,twocolumn,showpacs,preprintnumbers,
amsmath,amssymb,amsfonts,superscriptaddress]{revtex4}

\usepackage{graphicx}
\usepackage{dcolumn}
\usepackage{bm}
\usepackage{rotating}

\def\hermes{{\sc Hermes}}

\def\desy{{\sc Desy}}
\def\hera{{\sc Hera}}

\def\jetset{{\sc Jetset}}
\def\pythia6{{\sc Pythia6}}

\def\cmpi{\mbox{\large$\bm{\langle}$}\ensuremath{\sin(\phi+\phi_S)}
        \mbox{\large$\bm{\rangle}$}\ensuremath{_{UT}^\pi}}
\def\smpi{\mbox{\large$\bm{\langle}$}\ensuremath{\sin(\phi-\phi_S)}
        \mbox{\large$\bm{\rangle}$}\ensuremath{_{UT}^\pi}}
\def\cmh{\mbox{\large$\bm{\langle}$}\ensuremath{\sin(\phi+\phi_S)}
        \mbox{\large$\bm{\rangle}$}\ensuremath{_{UT}^h}}
\def\smh{\mbox{\large$\bm{\langle}$}\ensuremath{\sin(\phi-\phi_S)}
        \mbox{\large$\bm{\rangle}$}\ensuremath{_{UT}^h}}

\begin{document}


\def\groupalberta{\affiliation{Department of Physics, University of Alberta, Edmonton, Alberta T6G 2J1, Canada}}
\def\groupargonne{\affiliation{Physics Division, Argonne National Laboratory, Argonne, Illinois 60439-4843, USA}}
\def\groupbari{\affiliation{Istituto Nazionale di Fisica Nucleare, Sezione di Bari, 70124 Bari, Italy}}
\def\groupbeijing{\affiliation{School of Physics, Peking University, Beijing 100871, China}}
\def\groupchina{\affiliation{Department of Modern Physics, University of Science and Technology of China, Hefei, Anhui 230026, China}}
\def\groupcolorado{\affiliation{Nuclear Physics Laboratory, University of Colorado, Boulder, Colorado 80309-0446, USA}}
\def\groupdesy{\affiliation{\desy, 22603 Hamburg, Germany}}
\def\groupzeuthen{\affiliation{\desy\, 15738 Zeuthen, Germany}}
\def\groupdubna{\affiliation{Joint Institute for Nuclear Research, 141980 Dubna, Russia}}
\def\grouperlangen{\affiliation{Physikalisches Institut, Universit\"at Erlangen-N\"urnberg, 91058 Erlangen, Germany}}
\def\groupferrara{\affiliation{Istituto Nazionale di Fisica Nucleare, Sezione di Ferrara and Dipartimento di Fisica, Universit\`a di Ferrara, 44100 Ferrara, Italy}}
\def\groupfrascati{\affiliation{Istituto Nazionale di Fisica Nucleare, Laboratori Nazionali di Frascati, 00044 Frascati, Italy}}
\def\groupgent{\affiliation{Department of Subatomic and Radiation Physics, University of Gent, 9000 Gent, Belgium}}
\def\groupgiessen{\affiliation{Physikalisches Institut, Universit\"at Gie{\ss}en, 35392 Gie{\ss}en, Germany}}
\def\groupglasgow{\affiliation{Department of Physics and Astronomy, University of Glasgow, Glasgow G12 8QQ, United Kingdom}}
\def\groupillinois{\affiliation{Department of Physics, University of Illinois, Urbana, Illinois 61801-3080, USA}}
\def\groupmit{\affiliation{Laboratory for Nuclear Science, Massachusetts Institute of Technology, Cambridge, Massachusetts 02139, USA}}
\def\groupmichigan{\affiliation{Randall Laboratory of Physics, University of Michigan, Ann Arbor, Michigan 48109-1120, USA }}
\def\groupmoscow{\affiliation{Lebedev Physical Institute, 117924 Moscow, Russia}}
\def\groupnikhef{\affiliation{Nationaal Instituut voor Kernfysica en Hoge-Energiefysica (NIKHEF), 1009 DB Amsterdam, The Netherlands}}
\def\groupstpetersburg{\affiliation{Petersburg Nuclear Physics Institute, St. Petersburg, Gatchina, 188350 Russia}}
\def\groupprotvino{\affiliation{Institute for High Energy Physics, Protvino, Moscow region, 142281 Russia}}
\def\groupregensburg{\affiliation{Institut f\"ur Theoretische Physik, Universit\"at Regensburg, 93040 Regensburg, Germany}}
\def\grouprome{\affiliation{Istituto Nazionale di Fisica Nucleare, Sezione Roma 1, Gruppo Sanit\`a and Physics Laboratory, Istituto Superiore di Sanit\`a, 00161 Roma, Italy}}
\def\groupsimonfraser{\affiliation{Department of Physics, Simon Fraser University, Burnaby, British Columbia V5A 1S6, Canada}}
\def\grouptriumf{\affiliation{TRIUMF, Vancouver, British Columbia V6T 2A3, Canada}}
\def\grouptokyo{\affiliation{Department of Physics, Tokyo Institute of Technology, Tokyo 152, Japan}}
\def\groupamsterdam{\affiliation{Department of Physics and Astronomy, Vrije Universiteit, 1081 HV Amsterdam, The Netherlands}}
\def\groupwarsaw{\affiliation{Andrzej Soltan Institute for Nuclear Studies, 00-689 Warsaw, Poland}}
\def\groupyerevan{\affiliation{Yerevan Physics Institute, 375036 Yerevan, Armenia}}


\groupalberta
\groupargonne
\groupbari
\groupbeijing
\groupchina
\groupcolorado
\groupdesy
\groupzeuthen
\groupdubna
\grouperlangen
\groupferrara
\groupfrascati
\groupgent
\groupgiessen
\groupglasgow
\groupillinois
\groupmit
\groupmichigan
\groupmoscow
\groupnikhef
\groupstpetersburg
\groupprotvino
\groupregensburg
\grouprome
\groupsimonfraser
\grouptriumf
\grouptokyo
\groupamsterdam
\groupwarsaw
\groupyerevan


\author{A.~Airapetian}  \groupmichigan
\author{N.~Akopov}  \groupyerevan
\author{Z.~Akopov}  \groupyerevan
\author{M.~Amarian}  \groupzeuthen \groupyerevan
\author{A.~Andrus}  \groupillinois
\author{E.C.~Aschenauer}  \groupzeuthen
\author{W.~Augustyniak}  \groupwarsaw
\author{R.~Avakian}  \groupyerevan
\author{A.~Avetissian}  \groupyerevan
\author{E.~Avetissian}  \groupfrascati
% ...

\collaboration{The \hermes\ Collaboration} \noaffiliation

\title{Single-spin asymmetries in
semi-inclusive deep-inelastic scattering on a transversely polarized 
hydrogen target }

\date{\today}

\begin{abstract}
Single-spin asymmetries for semi-inclusive electroproduction of
charged pions in deep-inelastic scattering of
positrons are measured for the first time with transverse target
polarization.  The 
asymmetry depends on the azimuthal angles of both the pion
($\phi$) and the target spin axis ($\phi_S$) about the virtual
photon direction and relative to the lepton scattering plane. The
extracted Fourier component $\cmpi$
is a signal of the previously unmeasured quark transversity distribution, 
in conjunction with the so-called Collins fragmentation function, 
also unknown.  The Fourier component $\smpi$
of the asymmetry arises from a correlation
between the transverse polarization of the target nucleon and the
intrinsic transverse momentum of quarks, as represented by the 
previously unmeasured Sivers distribution function. 
Evidence for both signals is observed, but the Sivers asymmetry
may be affected by exclusive vector meson production.
\end{abstract}

\maketitle

The nucleon is a bound state containing quarks with momenta of order
$\Lambda_{QCD}\simeq 200$\,MeV.  As the masses of the
quarks of flavor $q=up (u)$ or $down (d)$ are much smaller than this, their
internal motion is relativistic.  
The quark substructure of hadrons is often probed in the Deep-Inelastic
Scattering (DIS) of leptons {\em i.e.}\ the absorption by a quark of a spacelike
virtual photon with large squared four-momentum  $q^2=-Q^2$.  The essence
of these experiments is captured
by the Parton Model in which partons (ie.
quarks and gluons) are scattered quasielastically by the lepton.  The
physics is most transparent in the frame in which the nucleon target
moves contrary  to the photon with ``infinite" momentum.
In this frame the transverse motion of the partons is ``frozen" 
during the interaction time, while their transverse momenta
are obviously unchanged.  After averaging over
this intrinsic transverse momentum $p_T$, three fundamental
distributions in longitudinal quark momentum can be interpreted
as number densities.  Two of these have been experimentally
explored in some detail---the unpolarized density
$q(x)$~\cite{pdf:MRST2001,pdf:cteq6}, and the helicity density
$\Delta q(x)\equiv q^{{\stackrel{\rightarrow}{\Rightarrow}}}(x)-
                                q^{\stackrel{\rightarrow}{\Leftarrow}}(x)$ 
reflecting the probability of finding the helicity of the quark to be
the same as that of the target nucleon~\cite{Lampe:1998eu}.
Here $x=Q^2/(2P\cdot q)$ is the dimensionless Bjorken scaling variable
representing the momentum fraction of the target nucleon carried by the
parton, where $P$ is the four-momentum of the target proton.
Viewed in the same helicity basis, the third distribution 
known as transversity~\cite{ph:RS,ph:ArtruM,ph:JJ},
$\delta q$ or alternatively $h_1^q$, is related to a 
forward scattering amplitude involving
helicity flip of both quark and target nucleon ($N^\mathbf{\Rightarrow} 
q^\leftarrow \rightarrow N^\mathbf{\Leftarrow}q^\rightarrow$) and 
has no probabilistic interpretation in this basis.  
However, it is a number
density in a basis of transverse spin eigenstates:  
$\delta q = q^{\uparrow\Uparrow} - q^{\uparrow\Downarrow}$. 
The transversity and helicity densities may differ because
dynamically bound light quarks move relativistically, in which regime boosts 
and rotations don't commute.  Hence the measurement of these differences
can shed light on the dynamics of nonperturbative QCD.

\begin{figure}
\begin{center}
\includegraphics[height=6cm,angle=-90]{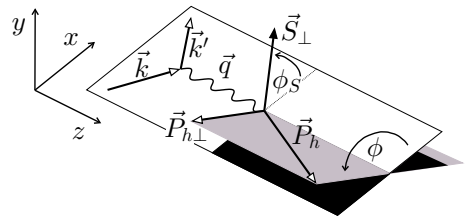}
\end{center}
\caption{\label{fig:phis} The definitions of the azimuthal
angles of the hadron production plane and the axis of the 
relevant component $\vec{S}_\perp$ of the target spin,
relative to the plane containing the momentum $\vec{k}$ ($\vec{k}'$) of the
incident (scattered) lepton.
Explicitly,
$\phi = \frac{\vec{q}\times\vec{k}\cdot\vec{P}_h}
             {|\vec{q}\times\vec{k}\cdot\vec{P}_h|}
\cos^{-1}{\frac{\vec{q}\times\vec{k}\cdot\vec{q}\times\vec{P}_h}
               {|\vec{q}\times\vec{k}||\vec{q}\times\vec{P}_h|}}$
and $\phi_S = \frac{\vec{q}\times\vec{k}\cdot\vec{S}_\perp}
                   {|\vec{q}\times\vec{k}\cdot\vec{S}_\perp|}
\cos^{-1} {\frac{\vec{q}\times\vec{k}\cdot\vec{q}\times\vec{S}_\perp}
                {|\vec{q}\times\vec{k}||\vec{q}\times\vec{S}_\perp|}}$,
where $0<\cos^{-1}<\pi$.
 }
\end{figure}
Transversity has thus far remained unmeasured because 
it is chiral-odd, and hard interactions conserve chirality.
However, it may be probed by a process involving some additional
chiral-odd structure.  If a hadron produced from the struck quark is
detected in addition to the scattered lepton, the transverse 
polarization of the struck quark can influence the 
transverse momentum component $\vec{P}_{h\perp}$ of the hadron
orthogonal to the virtual photon direction, and thereby influence 
its distribution in the azimuthal angle $\phi$ about 
the virtual photon direction relative to the lepton scattering 
plane~\cite{ph:Collins93} (see Fig.~\ref{fig:phis}.). The fragmentation 
function $H_{1}^\perp$ describing this spin-momentum correlation is 
indeed chiral-odd, and also odd under naive time reversal (T-odd),
which is time reversal without interchange of initial and final states. 
Known as the ``Collins function", it represents the interference 
of two amplitudes with different imaginary parts that can account for
single-spin asymmetries.
Such asymmetries involving longitudinal target polarization have
already been observed in pion 
electroproduction~\cite{hermes:azimUL}.
Theoretical 
interpretation~\cite{ph:BM00,ph:AM,ph:EGS,ph:MSJ,Efremov:2003eq,Schweitzer:2003yr} 
of those data in terms of transversity-related distributions, as well as 
theoretical calculations~\cite{ph:BKMM,Gamberg:2003eg}
suggest that the Collins function has a substantial magnitude, 
although the effects of pion and gluon rescattering
tend to cancel~\cite{Bacchetta:2003xn}.  Thus measurements 
employing transverse target polarization may be expected to
constrain transversity itself.  This Letter is the first report of
experimental semi-inclusive DIS asymmetries with transverse target polarization.

A completely different possible mechanism for producing target-related 
single-spin asymmetries has also been identified.  It was realized 
over a decade ago that such asymmetries might arise from correlations
between the transverse polarization of the target nucleon and the 
$p_T$ of quarks~\cite{Meng:1989cr,ph:Sivers,Anselmino:1999pw}.  
A vestige of that quark $p_T$ surviving both the photo-absorption 
and the ordinary fragmentation process can be inherited in the 
transverse momentum component $\vec{P}_{h\perp}$ of the produced hadron,
and can thereby influence its azimuthal distribution relative to
the target spin axis.
The ``Sivers distribution function" $f_{1T}^\perp$
describing the correlation of $p_T$ with target polarization
is related to a forward scattering amplitude involving
helicity-flip of only the target nucleon ($N^\mathbf{\Rightarrow} 
q \rightarrow N^\mathbf{\Leftarrow}q$), which must therefore
involve orbital angular momentum of the unpolarized 
quark~\cite{ph:BHS,Burkardt:2003yg}.
Recently this idea has found a reformulation touching on
fundamental issues in QCD.  It was realized that single-spin
asymmetries that can be attributed to such $p_T$-dependent parton
distributions can also be
understood in terms of a final-state interaction (FSI) via a soft
gluon~\cite{ph:BHS,ph:Collins02,ph:JY}. 
This FSI is a leading-order approximation for a gauge link 
that is necessary to restore color gauge invariance~\cite{ph:BJY}.
A key point is that the FSI offers a
mechanism for the interference of amplitudes that is
associated with the T-odd nature of the Sivers function,
which was once believed to forbid its existence.  
A related chiral-odd partner $h_1^\perp$~\cite{ph:BoerM} 
of the chiral-even Sivers
function was found to provide an explanation for the substantial
$\cos 2\phi$ dependence observed in Drell-Yan cross
sections~\cite{ph:Boer,ph:BBH}.  The Sivers function itself is
predicted to create Drell-Yan single-spin
asymmetries~\cite{ph:ADM}, but there it is believed to have the 
opposite sign to its appearance in DIS, due to the fundamental time
reversal symmetry of QCD~\cite{ph:Collins02}.  This prediction
of perturbative QCD needs to be tested experimentally.

Single-spin azimuthal asymmetries arising from the Collins and
Sivers mechanisms have a common $\sin\phi$ behavior when the
target is polarized along the lepton beam axis, as was the case
for all previously published single-spin asymmetries for leptoproduction.
However, the additional degree of
freedom that is the azimuthal angle $\phi_S$ of the axis of
transverse target polarization results in distinctive signatures:
$\sin(\phi-\phi_S)$ for the Sivers mechanism, and
$\sin(\phi+\phi_S)$ for the Collins mechanism~\cite{ph:BoerM}.  
Only the Collins mechanism involves the orientation of 
the lepton scattering plane because it depends on the
influence of the quark's polarization on that component of the 
transverse momentum $k_T$ acquired in the fragmentation process 
by the struck quark that is orthogonal to its transverse polarization,
{\em after} its spin component in the lepton scattering plane
has been flipped by the photo-absorption. 
In contrast, the Sivers effect arises through the struck quark
``remembering" the $p_T$ that it had in the target.
In either case, the transverse momentum tends to be
inherited by a forward hadron that may ``contain" this quark.
Hence the hadron $P_{h\perp}$ is correlated with
$k_T$ ($p_T$) in the case of the Collins (Sivers) effect.

In the analysis reported here, the cross section asymmetry with respect to the
target polarization is extracted as a two-dimensional distribution 
in $\phi$ versus $\phi_S$, which is then fitted 
with a sum of contributions from the above two sinusoidal dependences.
This simultaneous extraction of both 
contributions was shown by detailed Monte Carlo
simulations to avoid significant cross-contamination, even when they have
very different magnitudes in the context of a limited detector acceptance.  

The data reported here were recorded during the 2002--2003 running period of
the \hermes\ experiment using a transversely nuclear-polarized
hydrogen gas target internal to the $E\!=\!27.5$\,GeV \hera\ positron
storage ring at \desy.  The positron beam was unpolarized at this
time. The open-ended target cell is fed by an atomic-beam source
based on Stern-Gerlach separation~\cite{hermes:ABS} with
hyperfine transitions. The nuclear polarization of the atoms is
flipped at 60\,s time intervals, while both this polarization and
the atomic fraction inside the target cell are continuously
measured~\cite{hermes:BRPTGA}.
The average value of the
proton polarization $S_T$ was $0.78\pm 0.04$.  Tracking
corrections were applied for the deflections of the
scattered particles caused by the vertical 0.3\,T target holding
field, with little effect on the extracted asymmetries.

Scattered positrons and any coincident hadrons are detected by
the \hermes\ spectrometer~\cite{hermes:spectr}.
Its acceptance spans the scattering angle range $40<|\theta_y|<140$ mrad 
and $|\theta_x|<170$ mrad.  Hence the azimuthal
acceptance is segmented, but this was found in Monte Carlo
studies to have negligible effect on the Fourier components of interest.
Positrons are identified with an efficiency exceeding 98\% and a
hadron contamination of less than 1\% using an electromagnetic
calorimeter, a transition-radiation detector, a preshower
scintillation counter and a {\v C}erenkov detector. Charged pions
are identified using a dual-radiator ring-imaging {\v C}erenkov
detector~\cite{hermes:rich}.

Events were selected subject to the kinematic requirements $W^2 >
10$\,GeV$^2$, $0.1 < y < 0.85$ and $Q^2>1$\,GeV$^2$, where $W$ is the
invariant mass of the initial photon-nucleon system and
$y=(P\cdot q)/(P\cdot k)$.
Coincident hadrons were accepted
if $0.2<z<0.7$ and $\theta_{\gamma^* h} > 0.02$\,rad, 
where $z=(P\cdot P_h)/(P\cdot q)$, and
$\theta_{\gamma^* h}$ is
the angle between the directions of the virtual photon and the
hadron.  All hadrons detected in each event were included --- 
not only the one with largest $z$.  

For each produced hadron type $h$, and for each bin in either $x$
or $z$ or for the entire data set, the asymmetry was evaluated in
two dimensions $\phi$ and $\phi_S$, where $\phi_S$ always indicates
the spin direction of the $\mathbf{\uparrow}$ state.  Defining
$N_h^{\mathbf{\uparrow}(\mathbf{\downarrow})}(\phi,\phi_S)$
as the semi-inclusive luminosity-normalized yield in that target
spin state, the asymmetry is
\begin{equation} \label{eq:Aphi}
A^h_{UT}(\phi,\phi_S) = \frac{1}{|S_T|}
\frac{\left(N_h^\mathbf{\uparrow}(\phi,\phi_S) 
  - N_h^\mathbf{\downarrow}(\phi,\phi_S)\right)}
{\left(N_h^\mathbf{\uparrow}(\phi,\phi_S) 
  + N_h^\mathbf{\downarrow}(\phi,\phi_S)\right)}\,,
\end{equation}
The Collins azimuthal moment $\cmh$ and Sivers moment $\smh$
of the virtual-photon asymmetry are extracted in the fit
\begin{eqnarray} \nonumber
\frac{A^h_{UT}(\phi,\phi_S)}{2} &=& \cmh 
\frac{B(\langle y\rangle)}
     {A(\langle x\rangle,\langle y\rangle)}
\sin(\phi+\phi_S) \\
&+& \smh \sin(\phi-\phi_S)\,.
\end{eqnarray}
Here $B(y) \equiv (1-y)$,
$A(x,y) \equiv \frac{y^2}{2}+(1-y)(1+R(x,y))/(1+\gamma(x,y)^2)$,
$R(x,y)$ is the ratio of longitudinal to transverse DIS cross sections,
$\gamma(x,y)^2 \equiv 2 M_p x/(E y)$.  
The values for $R(\langle x\rangle,\langle y\rangle)$~\cite{pdf:R1990}
cannot be neglected here as they fall in the range 0.1--0.34.  
The reduced-$\chi^2$ values for the fits are in the range 0.74--1.89.
The statistical correlations
between the Sivers and Collins moments fall in the range -0.5 to -0.6.
The addition of terms for $\sin(3\phi-\phi_S)$, $\sin\phi_S$ and
$\sin(2\phi-\phi_S)$ resulted in coefficients that are negligible 
compared to their uncertainties, and in negligible changes
to the Collins and Sivers moments.
Effects of acceptance, instrumental smearing and QED radiation were all
found to be negligible in Monte Carlo simulations~\cite{hermes:radgen}.
The largest contribution to the systematic
uncertainties is due to the target polarization.

\begin{figure}[h]
\begin{center}
\includegraphics[width=85mm]{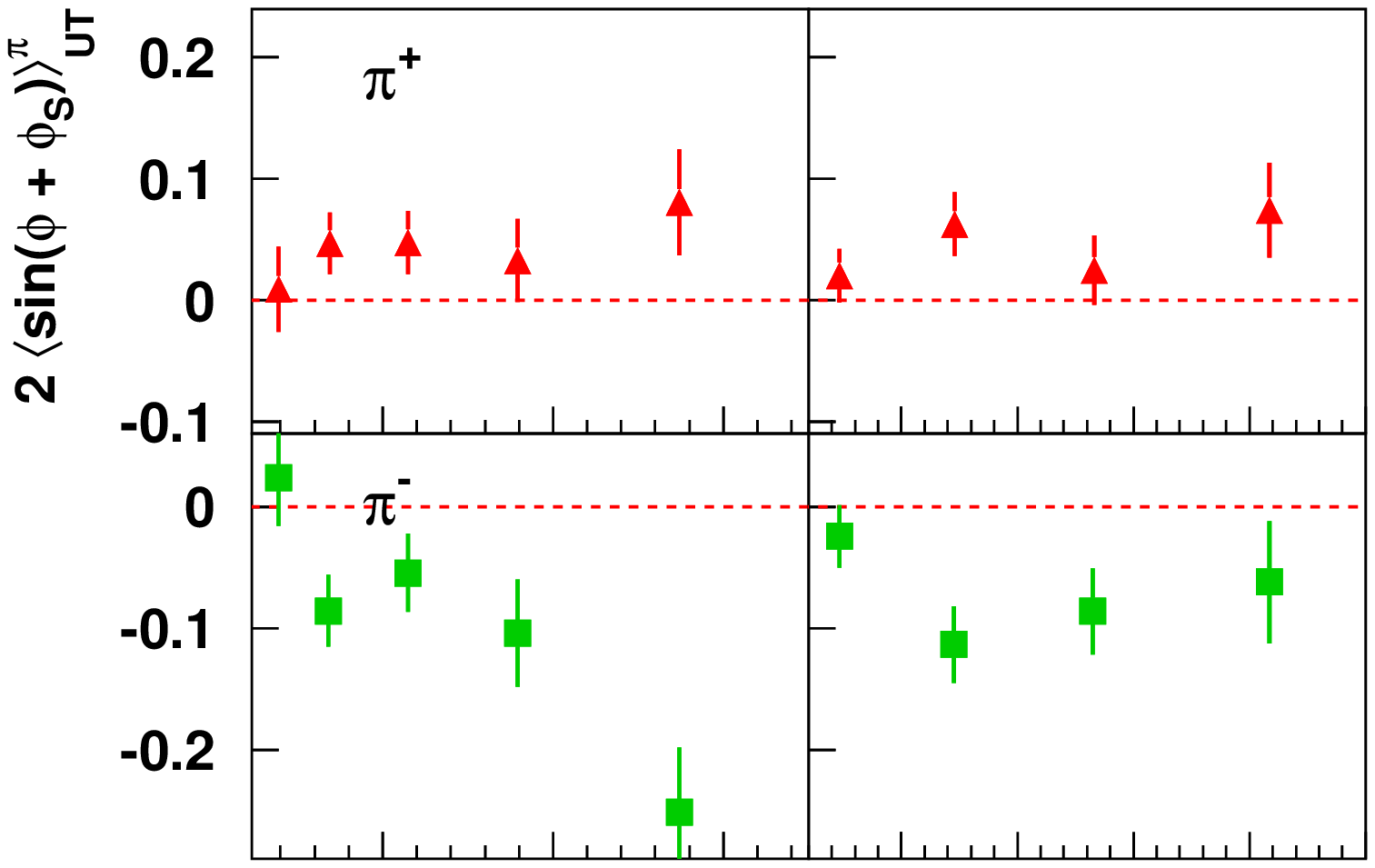} \hfill \\
\includegraphics[width=85mm]{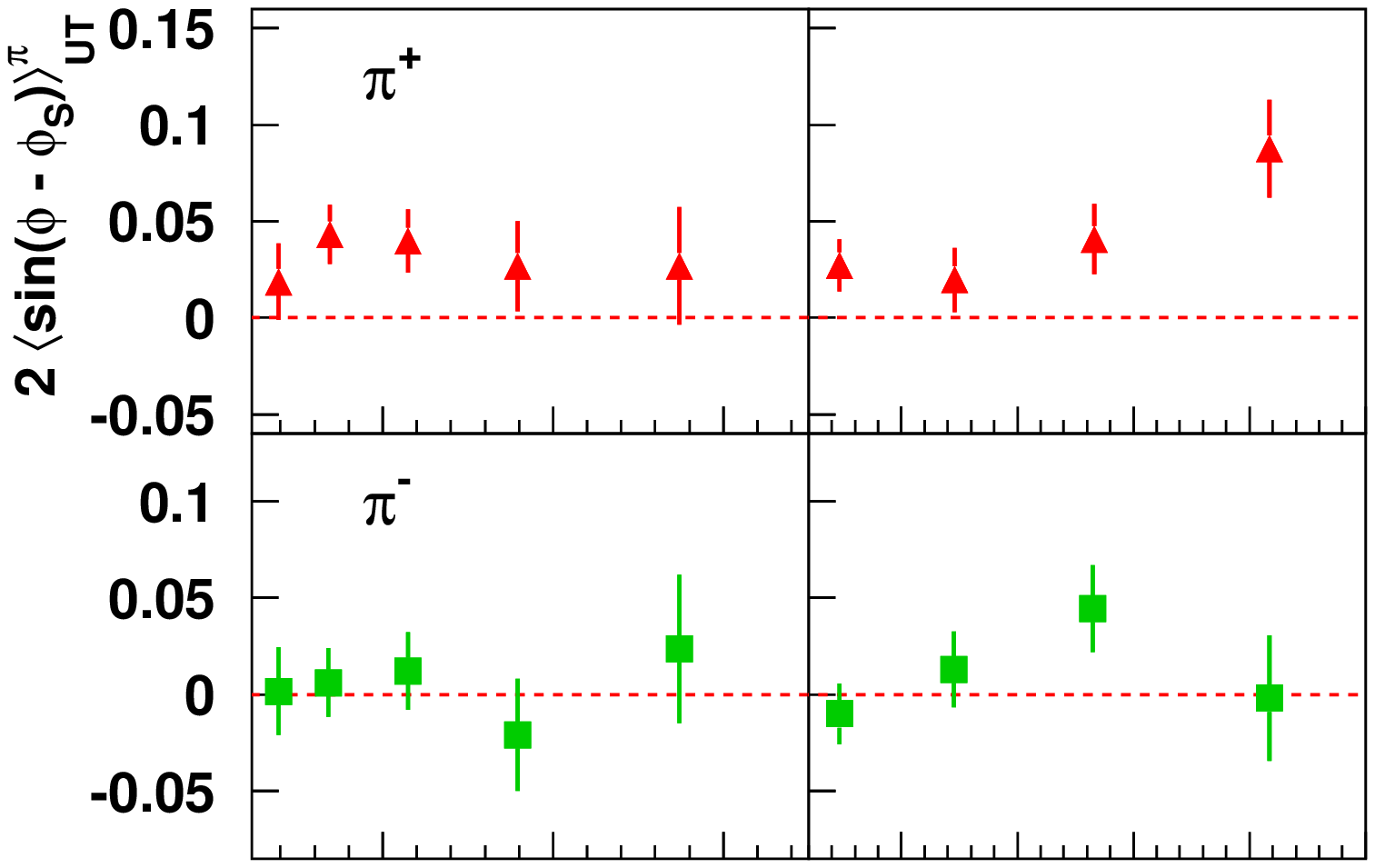} \hfill \\
\includegraphics[width=85mm]{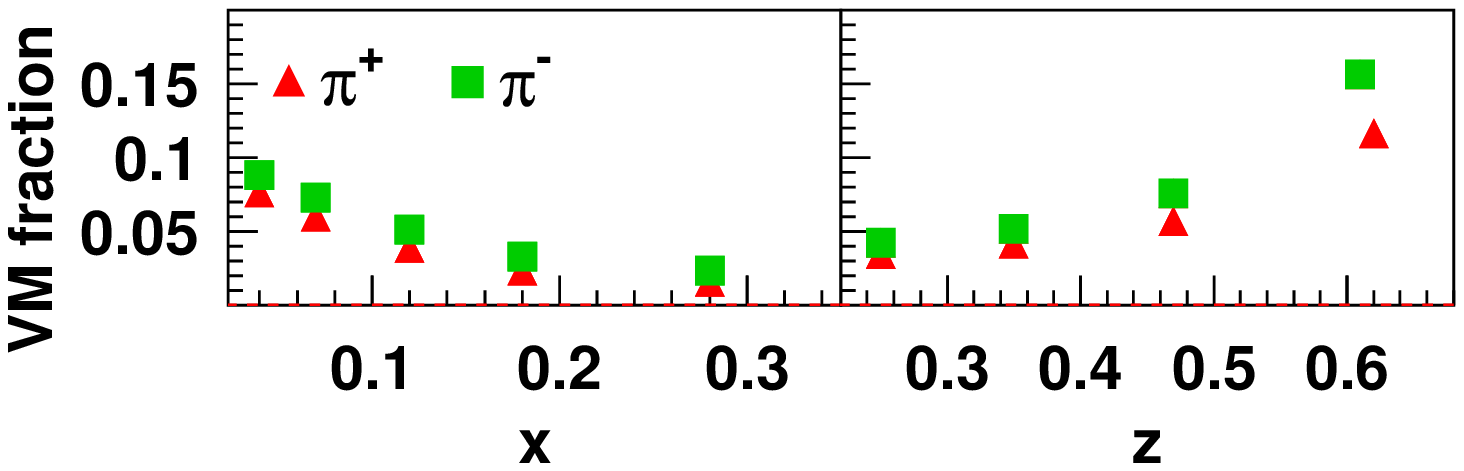}
\end{center}
\caption{\label{fig:asymm} Virtual-photon Collins (Sivers) moments
for charged pions as labelled in the upper (middle) panel, 
as a function of $x$ and $z$, multiplied by two to have the possible
range $\pm1$. 
The error bars represent the statistical uncertainties.
In addition, there is a common 8\% scale uncertainty in the moments.
The lower panel shows the relative contributions to the data from
simulated exclusive vector meson production.
 }
\end{figure}

When the azimuthal moments are averaged over the
experimental acceptance, 
the selected ranges in $x$ and $z$ are $0.023<x<0.4$ and $0.2<z<0.7$, 
and the corresponding mean values of the kinematic parameters are
$\langle x \rangle = 0.09, \langle y \rangle = 0.54, 
\langle Q^2 \rangle = 2.41$\,GeV$^2,
\langle z \rangle = 0.36 $ and $ \langle P_{\pi\perp} \rangle = 0.41$\,GeV.
The dependences of the charged pion moments on $x$ and $z$ are shown in
Fig.~\ref{fig:asymm}.  Also shown are simulations based on 
\pythia6~\cite{ph:jetset},
tuned for \hermes\ kinematics, of the fractions of the semi-inclusive
pion yield from exclusive production of vector mesons, the asymmetries
of which are poorly determined.

The averaged Collins moment for $\pi^+$
is positive at $0.021\pm0.007$(stat), while it is negative 
at $-0.038\pm0.008$(stat) for $\pi^-$.
Such a difference is expected if the
transversity densities resemble the helicity densities to the extent that
$\delta u$ is positive and $\delta d$ is negative and smaller in
magnitude, as models predict~\cite{Barone:2001sp}.  However, the magnitude of 
the negative $\pi^-$ moment appears to be at least as large as that for $\pi^+$.
The left panel
shows that this trend becomes more apparent as the magnitudes of 
these transverse moments increase at larger $x$ where valence 
quarks tend to dominate, as did the previously measured longitudinal
asymmetries.  However, the large negative $\pi^-$ 
moments might be considered unexpected as neither quark flavor 
dominates $\pi^-$ production like the $up$ quark dominates $\pi^+$, 
and one expects $|\delta d|<|\delta u|$ in analogy with
$|\Delta d|<|\Delta u|$.  This expectation is reflected in model 
predictions~\cite{Efremov:2003eq,Schweitzer:2003yr} based on 
the interpretation of those longitudinal asymmetries.  This
failure of those predictions could be due to the neglect of 
T-odd distributions such as the
Sivers function, the contribution of sea quarks or 
disfavored Collins fragmentation.  

One explanation of the larger negative $\pi^-$ azimuthal moments could be 
a substantial magnitude with opposite sign for the disfavored 
Collins function describing {\em e.g.} the fragmentation of 
$up$ quarks to $\pi^-$ mesons.  
Opposite signs of the favored and disfavored Collins
functions might be understood in the light of the 
string model of fragmentation. If a
favored pion forms as the string end created by
the first break, a disfavored pion from the next break will
inherit transverse momentum from the first break in the opposite
direction from that acquired by the first pion.  Such a $P_{\pi\perp}$ 
anticorrelation between favored and disfavored pions is demonstrated
by the \jetset\ simulation~\cite{ph:jetset}, which is based on a string
fragmentation model.  Hence any
correlation between $P_{\pi\perp}$ and another
kinematic or spin observable should have the opposite
sign for favored and disfavored pions.

The averaged Sivers moment is positive and nonzero 
at $0.017\pm0.004$(stat) for $\pi^+$,
appearing to provide the first evidence in leptoproduction 
for a T-odd parton distribution function.  
The $\pi^-$ moment is consistent with zero: $0.002\pm0.005$(stat).
Since the $\pi^+$ moment is dominated by $up$ quarks,
a positive value with the definition of azimuthal angles used here
would imply a negative value for the Sivers function of this flavor.
Continuing studies of the small sample 
of exclusive $\rho^0$ events in which both decay pions are detected suggest
that this asymmetry extracted for the $\pi^+$ exactly as 
in the semi-inclusive DIS analysis also has a significant 
positive tendency.  (The Collins asymmetries from this event 
sample are consistent with zero.)  
Therefore the Sivers asymmetry from the entire data 
set must be interpreted with caution.

In summary, a measurement with transverse target polarization of
single-spin asymmetries for semi-inclusive electroproduction of 
charged pions in deep-inelastic scattering has for the first time 
disentangled two
different phenomena that were indistinguishable in previous
data.  Their signals were extracted as distinctive Fourier components
of the dependence of the target-spin asymmetry on the azimuthal angles
of both the pion and the target spin axis about the virtual photon
direction and relative to the lepton scattering plane. One signal 
can arise from the transverse polarization of quarks in the
target, revealed by its influence on the fragmentation of the struck
quark.  A surprising feature of the data can be explained
by the hypothesis that fragmentation that is disfavored in terms of
quark flavor has an unexpected importance, and enters with a sign
opposite to that of the favored case.  The other
signal can arise from a correlation between the
transverse polarization of the target nucleon and the intrinsic
transverse momentum of quarks, and could provide another 
indication for nonzero orbital angular momentum of quarks 
in the nucleon.  A significant positive $\pi^+$ asymmetry has been
observed, corresponding to a negative value of the naively T-odd 
Sivers distribution function describing this correlation, 
but it may be due to a much larger apparent $\pi^+$ asymmetry in 
a small contamination from exclusive production of $\rho^0$ mesons.

\begin{acknowledgments} 
We gratefully acknowledge the \desy\ 
management for its support and the staff at \desy\ and the
collaborating institutions for their significant effort, and our
funding agencies for financial support.
\end{acknowledgments}

\bibliography{h1letter}

\begin{thebibliography}{37}
\expandafter\ifx\csname natexlab\endcsname\relax\def\natexlab#1{#1}\fi
\expandafter\ifx\csname bibnamefont\endcsname\relax
  \def\bibnamefont#1{#1}\fi
\expandafter\ifx\csname bibfnamefont\endcsname\relax
  \def\bibfnamefont#1{#1}\fi
\expandafter\ifx\csname citenamefont\endcsname\relax
  \def\citenamefont#1{#1}\fi
\expandafter\ifx\csname url\endcsname\relax
  \def\url#1{\texttt{#1}}\fi
\expandafter\ifx\csname urlprefix\endcsname\relax\def\urlprefix{URL }\fi
\providecommand{\bibinfo}[2]{#2}
\providecommand{\eprint}[2][]{\url{#2}}

\bibitem[{\citenamefont{Martin et~al.}(2002)}]{pdf:MRST2001}
\bibinfo{author}{\bibfnamefont{A.~D.} \bibnamefont{Martin}}
  \bibnamefont{et~al.}, \bibinfo{journal}{Eur. Phys. J.~C}
  \textbf{\bibinfo{volume}{23}}, \bibinfo{pages}{73} (\bibinfo{year}{2002}).

\bibitem[{\citenamefont{Pumplin et~al.}(2002)}]{pdf:cteq6}
\bibinfo{author}{\bibfnamefont{J.}~\bibnamefont{Pumplin}} \bibnamefont{et~al.},
  \bibinfo{journal}{JHEP} \textbf{\bibinfo{volume}{07}}, \bibinfo{pages}{012}
  (\bibinfo{year}{2002}), \eprint{hep-ph/0201195}.

\bibitem[{\citenamefont{Lampe and Reya}(2000)}]{Lampe:1998eu}
\bibinfo{author}{\bibfnamefont{B.}~\bibnamefont{Lampe}} \bibnamefont{and}
  \bibinfo{author}{\bibfnamefont{E.}~\bibnamefont{Reya}},
  \bibinfo{journal}{Phys. Rept.} \textbf{\bibinfo{volume}{332}},
  \bibinfo{pages}{1} (\bibinfo{year}{2000}).

\bibitem[{\citenamefont{Ralston and Soper}(1979)}]{ph:RS}
\bibinfo{author}{\bibfnamefont{J.~P.} \bibnamefont{Ralston}} \bibnamefont{and}
  \bibinfo{author}{\bibfnamefont{D.~E.} \bibnamefont{Soper}},
  \bibinfo{journal}{Nucl. Phys.} \textbf{\bibinfo{volume}{B152}},
  \bibinfo{pages}{109} (\bibinfo{year}{1979}).

\bibitem[{\citenamefont{Artru and Mekhfi}(1990)}]{ph:ArtruM}
\bibinfo{author}{\bibfnamefont{X.}~\bibnamefont{Artru}} \bibnamefont{and}
  \bibinfo{author}{\bibfnamefont{M.}~\bibnamefont{Mekhfi}},
  \bibinfo{journal}{Z. Phys. C} \textbf{\bibinfo{volume}{45}},
  \bibinfo{pages}{669} (\bibinfo{year}{1990}).

\bibitem[{\citenamefont{Jaffe and Ji}(1992)}]{ph:JJ}
\bibinfo{author}{\bibfnamefont{R.~L.} \bibnamefont{Jaffe}} \bibnamefont{and}
  \bibinfo{author}{\bibfnamefont{X.}~\bibnamefont{Ji}}, \bibinfo{journal}{Nucl.
  Phys.} \textbf{\bibinfo{volume}{B375}}, \bibinfo{pages}{527}
  (\bibinfo{year}{1992}).

\bibitem[{\citenamefont{Collins}(1993)}]{ph:Collins93}
\bibinfo{author}{\bibfnamefont{J.~C.} \bibnamefont{Collins}},
  \bibinfo{journal}{Nucl. Phys.} \textbf{\bibinfo{volume}{B396}},
  \bibinfo{pages}{161} (\bibinfo{year}{1993}).

\bibitem[{\citenamefont{Airapetian et~al.}()}]{hermes:azimUL}
\bibinfo{author}{\bibfnamefont{A.}~\bibnamefont{Airapetian}}
  \bibnamefont{et~al.} (\bibinfo{collaboration}{\hermes}),
  \bibinfo{howpublished}{Phys. Rev. Lett. {\bf 84}, 4047 (2002); Phys. Rev. D
  {\bf 64}, 097101 (2001); Phys. Lett. {\bf B562}, 182 (2003)}.

\bibitem[{\citenamefont{Boglione and Mulders}(2000)}]{ph:BM00}
\bibinfo{author}{\bibfnamefont{M.}~\bibnamefont{Boglione}} \bibnamefont{and}
  \bibinfo{author}{\bibfnamefont{P.~J.} \bibnamefont{Mulders}},
  \bibinfo{journal}{Phys. Lett.} \textbf{\bibinfo{volume}{B478}},
  \bibinfo{pages}{114} (\bibinfo{year}{2000}).

\bibitem[{\citenamefont{Anselmino and Murgia}(2000)}]{ph:AM}
\bibinfo{author}{\bibfnamefont{M.}~\bibnamefont{Anselmino}} \bibnamefont{and}
  \bibinfo{author}{\bibfnamefont{F.}~\bibnamefont{Murgia}},
  \bibinfo{journal}{Phys. Lett.} \textbf{\bibinfo{volume}{B483}},
  \bibinfo{pages}{74} (\bibinfo{year}{2000}).

\bibitem[{\citenamefont{Efremov et~al.}(2002)\citenamefont{Efremov, Goeke, and
  Schweitzer}}]{ph:EGS}
\bibinfo{author}{\bibfnamefont{A.~V.} \bibnamefont{Efremov}},
  \bibinfo{author}{\bibfnamefont{K.}~\bibnamefont{Goeke}}, \bibnamefont{and}
  \bibinfo{author}{\bibfnamefont{P.}~\bibnamefont{Schweitzer}},
  \bibinfo{journal}{Eur. Phys. J. C} \textbf{\bibinfo{volume}{24}},
  \bibinfo{pages}{407} (\bibinfo{year}{2002}).

\bibitem[{\citenamefont{Ma et~al.}(2002)\citenamefont{Ma, Schmidt, and
  Yang}}]{ph:MSJ}
\bibinfo{author}{\bibfnamefont{B.-Q.} \bibnamefont{Ma}},
  \bibinfo{author}{\bibfnamefont{I.}~\bibnamefont{Schmidt}}, \bibnamefont{and}
  \bibinfo{author}{\bibfnamefont{J.-J.} \bibnamefont{Yang}},
  \bibinfo{journal}{Phys. Rev. D} \textbf{\bibinfo{volume}{66}},
  \bibinfo{pages}{094001} (\bibinfo{year}{2002}).

\bibitem[{\citenamefont{Efremov et~al.}(2003)\citenamefont{Efremov, Goeke, and
  Schweitzer}}]{Efremov:2003eq}
\bibinfo{author}{\bibfnamefont{A.~V.} \bibnamefont{Efremov}},
  \bibinfo{author}{\bibfnamefont{K.}~\bibnamefont{Goeke}}, \bibnamefont{and}
  \bibinfo{author}{\bibfnamefont{P.}~\bibnamefont{Schweitzer}},
  \bibinfo{journal}{Eur. Phys. J. C} \textbf{\bibinfo{volume}{32}},
  \bibinfo{pages}{337} (\bibinfo{year}{2003}).

\bibitem[{\citenamefont{Schweitzer and Bacchetta}(2004)}]{Schweitzer:2003yr}
\bibinfo{author}{\bibfnamefont{P.}~\bibnamefont{Schweitzer}} \bibnamefont{and}
  \bibinfo{author}{\bibfnamefont{A.}~\bibnamefont{Bacchetta}},
  \bibinfo{journal}{Nucl. Phys.} \textbf{\bibinfo{volume}{A732}},
  \bibinfo{pages}{106} (\bibinfo{year}{2004}).

\bibitem[{\citenamefont{Bacchetta et~al.}(2002)\citenamefont{Bacchetta, Kundu,
  Metz, and Mulders}}]{ph:BKMM}
\bibinfo{author}{\bibfnamefont{A.}~\bibnamefont{Bacchetta}},
  \bibinfo{author}{\bibfnamefont{R.}~\bibnamefont{Kundu}},
  \bibinfo{author}{\bibfnamefont{A.}~\bibnamefont{Metz}}, \bibnamefont{and}
  \bibinfo{author}{\bibfnamefont{P.~J.} \bibnamefont{Mulders}},
  \bibinfo{journal}{Phys. Rev.} \textbf{\bibinfo{volume}{D65}},
  \bibinfo{pages}{094021} (\bibinfo{year}{2002}).

\bibitem[{\citenamefont{Gamberg et~al.}(2003)\citenamefont{Gamberg, Goldstein,
  and Oganessyan}}]{Gamberg:2003eg}
\bibinfo{author}{\bibfnamefont{L.~P.} \bibnamefont{Gamberg}},
  \bibinfo{author}{\bibfnamefont{G.~R.} \bibnamefont{Goldstein}},
  \bibnamefont{and} \bibinfo{author}{\bibfnamefont{K.~A.}
  \bibnamefont{Oganessyan}}, \bibinfo{journal}{Phys. Rev. D}
  \textbf{\bibinfo{volume}{68}}, \bibinfo{pages}{051501}
  (\bibinfo{year}{2003}).

\bibitem[{\citenamefont{Bacchetta et~al.}(2003)\citenamefont{Bacchetta, Metz,
  and Yang}}]{Bacchetta:2003xn}
\bibinfo{author}{\bibfnamefont{A.}~\bibnamefont{Bacchetta}},
  \bibinfo{author}{\bibfnamefont{A.}~\bibnamefont{Metz}}, \bibnamefont{and}
  \bibinfo{author}{\bibfnamefont{J.-J.} \bibnamefont{Yang}},
  \bibinfo{journal}{Phys. Lett.} \textbf{\bibinfo{volume}{B574}},
  \bibinfo{pages}{225} (\bibinfo{year}{2003}).

\bibitem[{\citenamefont{Meng et~al.}(1989)\citenamefont{Meng, Pan, Xie, and
  Zhu}}]{Meng:1989cr}
\bibinfo{author}{\bibfnamefont{T.}~\bibnamefont{Meng}},
  \bibinfo{author}{\bibfnamefont{J.}~\bibnamefont{Pan}},
  \bibinfo{author}{\bibfnamefont{Q.}~\bibnamefont{Xie}}, \bibnamefont{and}
  \bibinfo{author}{\bibfnamefont{W.}~\bibnamefont{Zhu}},
  \bibinfo{journal}{Phys. Rev. D} \textbf{\bibinfo{volume}{40}},
  \bibinfo{pages}{769} (\bibinfo{year}{1989}).

\bibitem[{\citenamefont{Sivers}(1990)}]{ph:Sivers}
\bibinfo{author}{\bibfnamefont{D.~W.} \bibnamefont{Sivers}},
  \bibinfo{journal}{Phys. Rev.~D} \textbf{\bibinfo{volume}{41}},
  \bibinfo{pages}{83} (\bibinfo{year}{1990}).

\bibitem[{\citenamefont{Anselmino et~al.}(1999)\citenamefont{Anselmino,
  Boglione, and Murgia}}]{Anselmino:1999pw}
\bibinfo{author}{\bibfnamefont{M.}~\bibnamefont{Anselmino}},
  \bibinfo{author}{\bibfnamefont{M.}~\bibnamefont{Boglione}}, \bibnamefont{and}
  \bibinfo{author}{\bibfnamefont{F.}~\bibnamefont{Murgia}},
  \bibinfo{journal}{Phys. Rev. D} \textbf{\bibinfo{volume}{60}},
  \bibinfo{pages}{054027} (\bibinfo{year}{1999}).

\bibitem[{\citenamefont{Brodsky et~al.}(2002)\citenamefont{Brodsky, Hwang, and
  Schmidt}}]{ph:BHS}
\bibinfo{author}{\bibfnamefont{S.~J.} \bibnamefont{Brodsky}},
  \bibinfo{author}{\bibfnamefont{D.~S.} \bibnamefont{Hwang}}, \bibnamefont{and}
  \bibinfo{author}{\bibfnamefont{I.}~\bibnamefont{Schmidt}},
  \bibinfo{journal}{Phys. Lett.} \textbf{\bibinfo{volume}{B530}},
  \bibinfo{pages}{99} (\bibinfo{year}{2002}).

\bibitem[{\citenamefont{Burkardt}(2004)}]{Burkardt:2003yg}
\bibinfo{author}{\bibfnamefont{M.}~\bibnamefont{Burkardt}},
  \bibinfo{journal}{Phys. Rev. D} \textbf{\bibinfo{volume}{69}},
  \bibinfo{pages}{057501} (\bibinfo{year}{2004}).

\bibitem[{\citenamefont{Collins}(2002)}]{ph:Collins02}
\bibinfo{author}{\bibfnamefont{J.~C.} \bibnamefont{Collins}},
  \bibinfo{journal}{Phys. Lett.} \textbf{\bibinfo{volume}{B536}},
  \bibinfo{pages}{43} (\bibinfo{year}{2002}).

\bibitem[{\citenamefont{Ji and Yuan}(2002)}]{ph:JY}
\bibinfo{author}{\bibfnamefont{X.}~\bibnamefont{Ji}} \bibnamefont{and}
  \bibinfo{author}{\bibfnamefont{F.}~\bibnamefont{Yuan}},
  \bibinfo{journal}{Phys. Lett.} \textbf{\bibinfo{volume}{B543}},
  \bibinfo{pages}{66} (\bibinfo{year}{2002}).

\bibitem[{\citenamefont{Belitsky et~al.}(2003)\citenamefont{Belitsky, Ji, and
  Yuan}}]{ph:BJY}
\bibinfo{author}{\bibfnamefont{A.~V.} \bibnamefont{Belitsky}},
  \bibinfo{author}{\bibfnamefont{X.}~\bibnamefont{Ji}}, \bibnamefont{and}
  \bibinfo{author}{\bibfnamefont{F.}~\bibnamefont{Yuan}},
  \bibinfo{journal}{Nucl. Phys.} \textbf{\bibinfo{volume}{B656}},
  \bibinfo{pages}{165} (\bibinfo{year}{2003}).

\bibitem[{\citenamefont{Boer and Mulders}(1998)}]{ph:BoerM}
\bibinfo{author}{\bibfnamefont{D.}~\bibnamefont{Boer}} \bibnamefont{and}
  \bibinfo{author}{\bibfnamefont{P.~J.} \bibnamefont{Mulders}},
  \bibinfo{journal}{Phys.~Rev.~D} \textbf{\bibinfo{volume}{57}},
  \bibinfo{pages}{5780} (\bibinfo{year}{1998}).

\bibitem[{\citenamefont{Boer}(1999)}]{ph:Boer}
\bibinfo{author}{\bibfnamefont{D.}~\bibnamefont{Boer}},
  \bibinfo{journal}{Phys.~Rev.~D} \textbf{\bibinfo{volume}{60}},
  \bibinfo{pages}{014012} (\bibinfo{year}{1999}).

\bibitem[{\citenamefont{Boer et~al.}(2003)\citenamefont{Boer, Brodsky, and
  Hwang}}]{ph:BBH}
\bibinfo{author}{\bibfnamefont{D.}~\bibnamefont{Boer}},
  \bibinfo{author}{\bibfnamefont{S.~J.} \bibnamefont{Brodsky}},
  \bibnamefont{and} \bibinfo{author}{\bibfnamefont{D.~S.} \bibnamefont{Hwang}},
  \bibinfo{journal}{Phys.~Rev.~D} \textbf{\bibinfo{volume}{67}},
  \bibinfo{pages}{054003} (\bibinfo{year}{2003}).

\bibitem[{\citenamefont{Anselmino et~al.}(2003)\citenamefont{Anselmino,
  D'Alesio, and Murgia}}]{ph:ADM}
\bibinfo{author}{\bibfnamefont{M.}~\bibnamefont{Anselmino}},
  \bibinfo{author}{\bibfnamefont{U.}~\bibnamefont{D'Alesio}}, \bibnamefont{and}
  \bibinfo{author}{\bibfnamefont{F.}~\bibnamefont{Murgia}},
  \bibinfo{journal}{Phys. Rev.~D} \textbf{\bibinfo{volume}{67}},
  \bibinfo{pages}{074010} (\bibinfo{year}{2003}).

\bibitem[{\citenamefont{Stock et~al.}(1994)}]{hermes:ABS}
\bibinfo{author}{\bibfnamefont{F.}~\bibnamefont{Stock}} \bibnamefont{et~al.},
  \bibinfo{journal}{Nucl. Inst. \& Meth.} \textbf{\bibinfo{volume}{A 343}},
  \bibinfo{pages}{334} (\bibinfo{year}{1994}).

\bibitem[{\citenamefont{Baumgarten et~al.}()}]{hermes:BRPTGA}
\bibinfo{author}{\bibfnamefont{C.}~\bibnamefont{Baumgarten}}
  \bibnamefont{et~al.}, \bibinfo{howpublished}{Nucl. Inst. \& Meth. {\bf A
  482}, 606 (2002); Nucl. Inst. \& Meth. {\bf A 496}, 263 (2003)}.

\bibitem[{\citenamefont{Ackerstaff et~al.}(1998)}]{hermes:spectr}
\bibinfo{author}{\bibfnamefont{K.}~\bibnamefont{Ackerstaff}}
  \bibnamefont{et~al.} (\bibinfo{collaboration}{\hermes}),
  \bibinfo{journal}{Nucl. Inst. \& Meth.} \textbf{\bibinfo{volume}{A 417}},
  \bibinfo{pages}{230} (\bibinfo{year}{1998}).

\bibitem[{\citenamefont{Akopov et~al.}(2002)}]{hermes:rich}
\bibinfo{author}{\bibfnamefont{N.}~\bibnamefont{Akopov}} \bibnamefont{et~al.},
  \bibinfo{journal}{Nucl. Inst. \& Meth.} \textbf{\bibinfo{volume}{A 479}},
  \bibinfo{pages}{511} (\bibinfo{year}{2002}).

\bibitem[{\citenamefont{Whitlow et~al.}(1990)}]{pdf:R1990}
\bibinfo{author}{\bibfnamefont{L.~W.} \bibnamefont{Whitlow}}
  \bibnamefont{et~al.}, \bibinfo{journal}{Phys. Lett.}
  \textbf{\bibinfo{volume}{B250}}, \bibinfo{pages}{193} (\bibinfo{year}{1990}).

\bibitem[{\citenamefont{Akushevich et~al.}(1998)}]{hermes:radgen}
\bibinfo{author}{\bibfnamefont{I.}~\bibnamefont{Akushevich}}
  \bibnamefont{et~al.} (\bibinfo{year}{1998}),
  \eprint[http://arXiv.org/abs]{hep-ph/9906408}.

\bibitem[{\citenamefont{Sj{\"o}strand et~al.}(2001)}]{ph:jetset}
\bibinfo{author}{\bibfnamefont{T.}~\bibnamefont{Sj{\"o}strand}}
  \bibnamefont{et~al.}, \bibinfo{journal}{Comput. Phys. Commun.}
  \textbf{\bibinfo{volume}{135}}, \bibinfo{pages}{238} (\bibinfo{year}{2001}).

\bibitem[{\citenamefont{Barone et~al.}(2002)\citenamefont{Barone, Drago, and
  Ratcliffe}}]{Barone:2001sp}
\bibinfo{author}{\bibfnamefont{V.}~\bibnamefont{Barone}},
  \bibinfo{author}{\bibfnamefont{A.}~\bibnamefont{Drago}}, \bibnamefont{and}
  \bibinfo{author}{\bibfnamefont{P.~G.} \bibnamefont{Ratcliffe}},
  \bibinfo{journal}{Phys. Rept.} \textbf{\bibinfo{volume}{359}},
  \bibinfo{pages}{1} (\bibinfo{year}{2002}).

\end{thebibliography}

\end{document}